\documentclass{article}
\usepackage{spconf,amsmath,graphicx,cite,epstopdf}

\title{Neural Kalman Filtering for Speech Enhancement}
\name{Wei Xue,  Gang Quan,  Chao Zhang, Guohong Ding, Xiaodong He, Bowen Zhou \thanks{This work was supported by JD-BAAI joint project.}}
\address{JD AI Research}
\begin{document}
%
\maketitle
\begin{abstract}
Conventional learning-based speech enhancement methods usually utilize existing building blocks to design the deep neural networks (DNNs), while how to effectively integrate the statistical signal processing based schemes, which are expert-knowledge driven and could ameliorate the over-fitting problem, into the network design remains an open issue. In this paper, we extend the conventional Kalman filtering (KF) and propose a supervised-learning based neural Kalman filter (NKF) for speech enhancement. Similar to KF, the proposed method first obtains a prediction from the speech evolution model and then integrates the short-term instantaneous observation by linear weighting, and the weights are calculated by comparing between the speech prediction residual error and the environmental noise level. An end-to-end network is designed to convert the speech linear prediction model in KF to non-linear, and to compact all other conventional linear filtering operations. Different with other DNN based methods, the proposed method provides a specialized network design inspired from the conventional signal processing, the backpropagation can be directly applied on the linear filtering operations integrated from KF. We conduct experiments in different noisy conditions, and the results demonstrate that the proposed method outperforms the baseline methods which are based on either signal processing or DNNs.
\end{abstract}
\section{Introduction}
Speech enhancement aims to suppress the environmental noise without distorting the target speech, and has wide applications in systems such as communication, hearing aids and automatic speech recognition.

Compared with the conventional signal processing based methods which use simple linear statistic models for speech and noise, such as Wiener filtering (WF) \cite{Wiener1949,Lim1979,Chen2006b}, subspace estimation \cite{Hansen1997,Hu2002,Ephraim1993} and Kalman filtering (KF) \cite{So2011a,Wang2018,Xue2018,Xue2018a,Xue2018b}, the speech enhancement performance has been dramatically improved by deep neural networks (DNNs) which learn the non-linear mapping from the noisy feature to the clean target. Generally, the networks are designed based on either the classical building blocks for DNN such as feed-forward network (FNN) \cite{Xu2014,Xu2015,Tu2017}, convolutional neural network (CNN) \cite{Mamun2019,Ouyang2019,Park2017}, recurrent neural network(RNN) \cite{Weninger2015,Gao2018,Sun2017}, or concatenations of these building blocks such as UNet\cite{Pandey2018,Yang2020,Choi2018} and convolutional recurrent neural network (CRNN) \cite{Tan2018,Strake2020}. Although the architectures are designed to effectively model the different time-frequency dependencies of speech and noise, there always lacks an explicit criterion for the model design, which makes it hard to interpret and optimize the intermediate representations, and also makes the performance highly rely on the diversity of the training data. In addition, many achievements in conventional signal processing based methods, which integrate the expert knowledge to derive the optimization steps and optimal filters given the statistics of speech and noise, have not fully been exploited by the supervised-learning based speech enhancement.

This paper proposes a neural Kalman filter (NKF) to fully integrate the statistical signal processing into DNN, by extending the KF to the supervised learning scheme. We note that \cite{Yu2019} uses a DNN to obtain a preprocessed speech for linear prediction (LP) model estimation in KF but uses the conventional KF for speech enhancement. An end-to-end network is designed in this paper to convert the speech LP model as well as the noise estimation model in KF to non-linear, and to compact all other conventional linear filtering operations. Clean speech estimates from RNN and WF are obtained, and are linearly combined by an NKF gain to yield the NKF output. The signal processing operations can serve as regularization for the network to avoid overfitting, and the backpropagation (BP) can be straightforwardly applied to the linear filtering processes. Moreover, the proposed method also overcomes the problem of unrealistic linear model assumptions in KF. We conduct experiments in different noisy conditions, and evaluation results demonstrate the effectiveness of the proposed method.

\section{Signal Model and KF}
\label{sec:kf}
We present the modulation-domain KF \cite{So2011a,Wang2018},  since it has shown superior performances than both time-domain and short-time Fourier transform (STFT)-domain  KFs. The modulation-domain KF regards the amplitude of speech in each frequency bin as a time-varying signal, and adopts a signal model as:
\begin{align}
|Y(t,f)| = |X(t,f)|+|V(t,f)|,
\label{sigmodel}
\end{align}
where $Y(t,f)$, $X(t,f)$, and $V(t,f)$ represent the STFT signals of the noisy speech, clean speech and noise, respectively, and $|\cdot|$ takes the amplitude. The clean speech amplitude $|X(t,f)|$ can be further expressed by a $P$-order LP model as:
\begin{align}
\mathbf{x}(t,f) = \mathbf{A}(f)\mathbf{x}(t-1,f) +\mathbf{u}W(t,f),
\label{eq2_lpc}
\end{align}
where $\mathbf{x}(t,f) = [|X(t,f), |X(t-1,f)|, ..., |X(t-P+1,f)|]^T$ is the hidden state vector, $\mathbf{A}(f)$ is the state transmission matrix defined in \cite{So2011a} according to the LP coefficients of speech, $\mathbf{u}= [1,0,...,0]^T$ is a  $P\times1$ vector, and $W(t,f)$ is the LP residual. In practice, the unknown LP coefficients are estimated via LP analysis on the output of WF.

\begin{figure}[t]
  \centering
  \includegraphics[width=60mm]{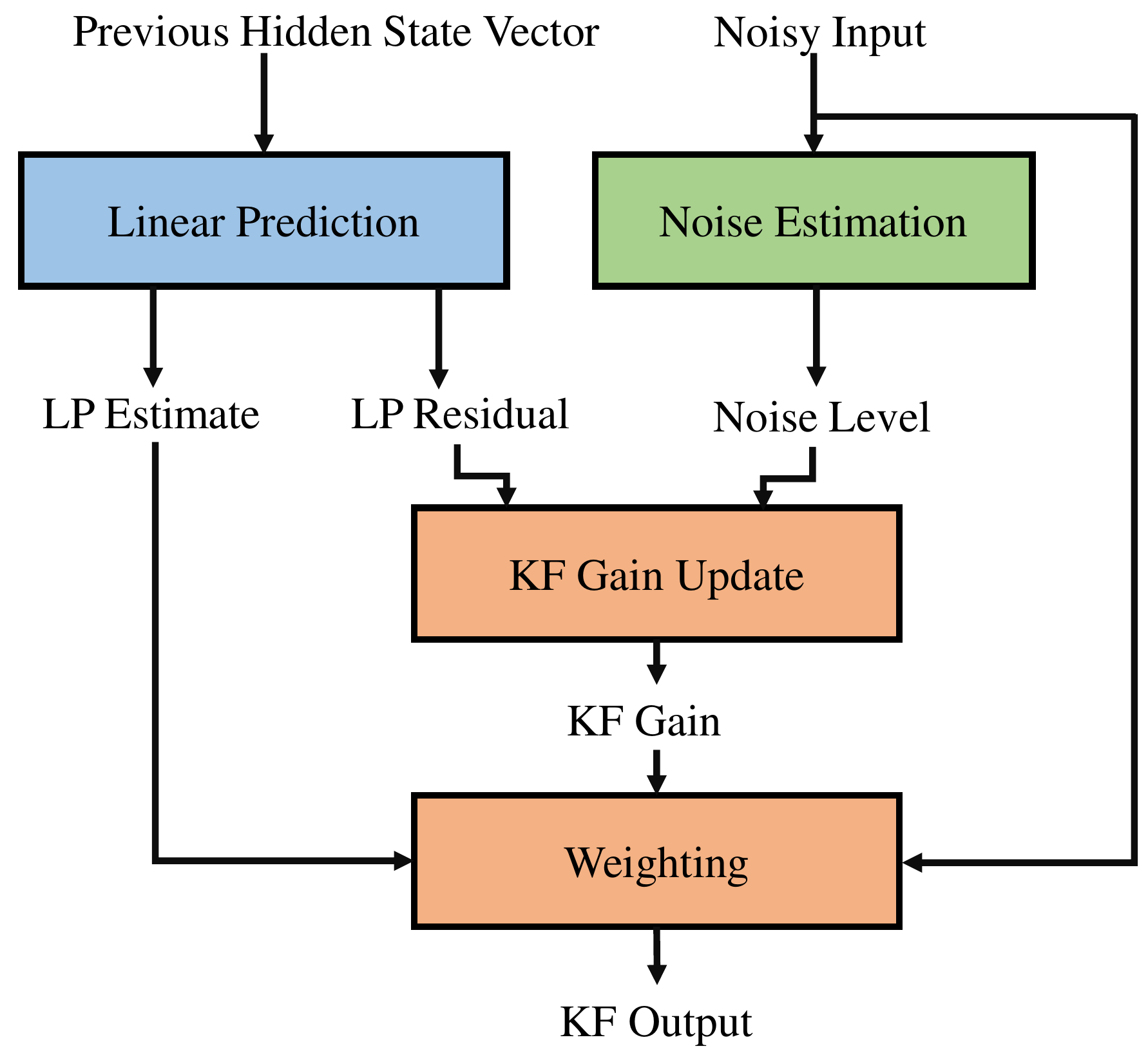}
  \caption{Diagram of KF based speech enhancement.}
  \label{fig:kf}
\end{figure}

As shown in Fig.~\ref{fig:kf}, the KF takes two stages for speech enhancement: predicting and updating.
Given the hidden state $\mathbf{x}(t-1|t-1,f)$ that consists of the clean speech estimates in the previous frame, the LP estimation of clean speech  $\mathbf{x}(t|t-1,f)$ is first obtained using \eqref{eq2_lpc} as:
\begin{align}
\mathbf{x}(t|t-1,f) = \mathbf{A}(f)\mathbf{x}(t-1|t-1,f),
\end{align}
and then used to update the KF estimate by incorporating the noisy observation in the current frame:
\begin{align}
&\hat{\mathbf{x}}(t|t,f)   \nonumber\\
&=[\mathbf{I}-\mathbf{G}(t,f)\mathbf{u}^T]\hat{\mathbf{x}}(t|t-1,f) + \mathbf{G}(t,f)|Y(t,f)|,
\label{kfeq4}
\end{align}
where the $\mathbf{G}(t,f)$ is the KF gain determined by comparing between the noise variance $\sigma^2_v = E\{V(t,f)V^*(t,f)\}$ and the variance matrix of the LP residual $\mathbf{R}_{ee}(t|t-1,f)$, as
\begin{align}
\mathbf{G}(t,f) = \frac{\mathbf{R}_{ee}(t|t-1,f)\mathbf{u}}{\sigma^2_v+\mathbf{u}^T\mathbf{R}_{ee}(t|t-1,f)\mathbf{u}}.
\label{kfeq5}
\end{align}
We observe that \eqref{kfeq5} takes a similar form to WF. It is also shown in \cite{Xue2018,Xue2018a,Xue2018b} that KF can be seen as introducing the speech evolution into WF.

\section{Proposed Method}

The KF gain in \eqref{kfeq4} is important to KF since it serves as a weighting factor such that the KF output adaptively approximates the LP estimate in noise-dominated TF bins, and approximates the instantaneous observation when the noise level is low. It is also worth noting that, essentially the output is a combination of $\hat{\mathbf{x}}(t|t-1,f)$ and $|Y(t,f)|$, while how to determine the weight remains a problem. The KF introduces the {\em LP residual} to select the KF gain. With the LP residual, it is shown that the KF gain can be optimally determined and updated under the minimum mean-squared error (MMSE) criterion, based on the statistics of LP estimate and noise.

\begin{figure}[t]
  \centering
  \includegraphics[width=80mm]{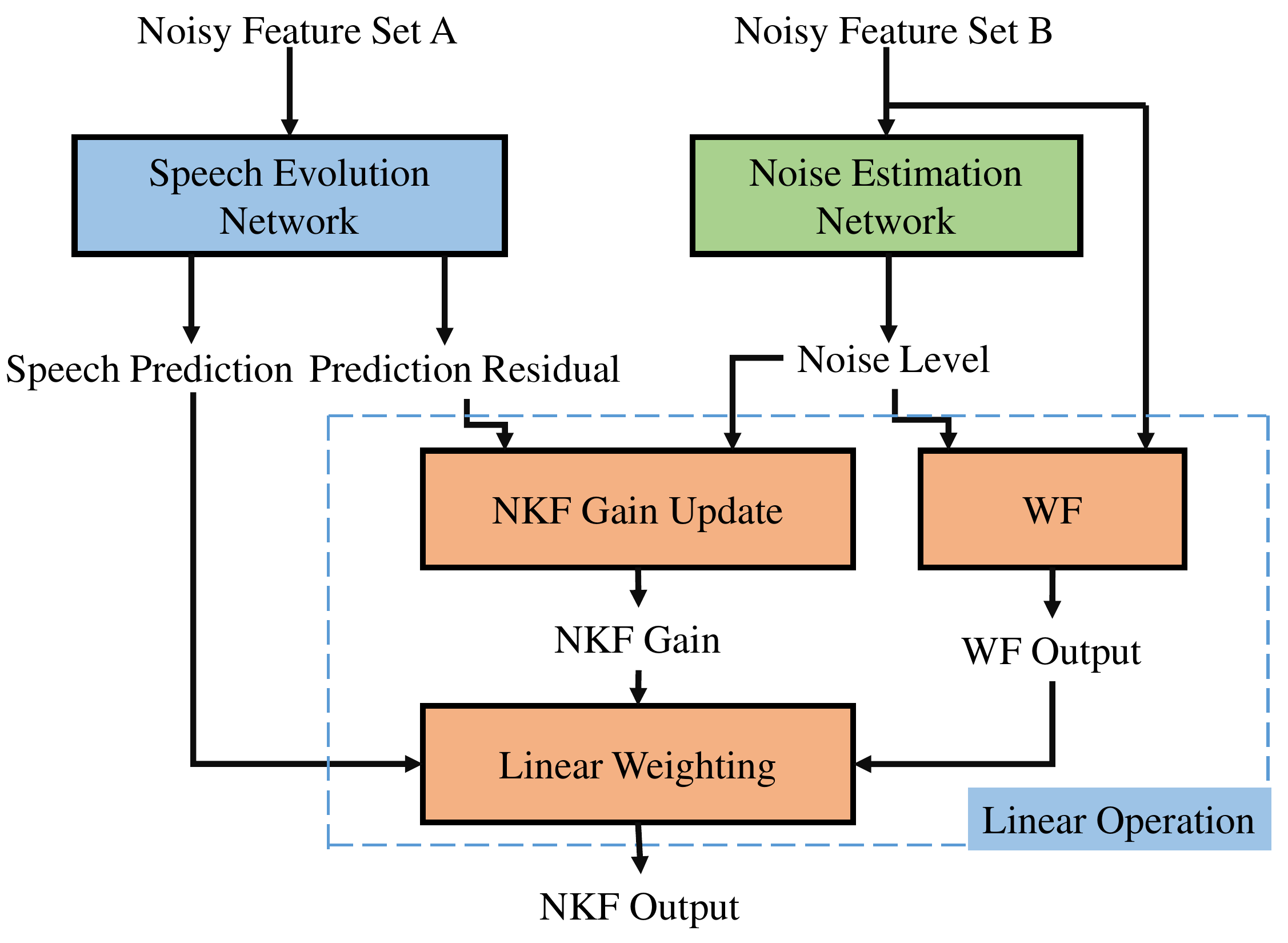}
  \caption{Diagram of the proposed NKF.}
  \label{fig:nkf}
\end{figure}

The above discussion motivates an integration of signal processing into the learning-based speech enhancement. Although networks have been proposed which generally contain modules with different expected functions, they usually rely on a large amount of diversified data to obtain the optimal network weights.  However, by using the expert knowledge to obtain the optimal solutions from conventional signal processing, it is possible to impose constraints or define the mathematical formulations for the weights of the connections between different modules. In this way, the signal processing operations can be seen as the regularization for the network, which is helpful to reduce the reliance on the training data and cope with the over-fitting problem. In addition, since the signal processing operations are usually linear and differentiable, the network that integrates the signal processing operations can be straightforwardly optimized using the BP algorithm.

In this section, we extend the above KF to the supervised learning scheme and propose an NKF, whose structure is shown in  Fig.~\ref{fig:nkf}, and the linear operations without learnable parameters are highlighted. The proposed NKF takes a similar form to the conventional KF, which first predicts the clean speech from a speech evolution model, and then updates the estimation by incorporating short-term information from the observation. Different from the KF, neural networks are used to replace the LP model of speech with a more realistic non-linear model which is learned from data, and also integrates the noise estimation process into the network.  We note that the proposed NKF combines the network prediction with the WF output rather than the raw observation to obtain the final output,  for which the reason will be explained later. The analytical expression of the NKF gain is defined according to the optimal KF gain derived under the MMSE criterion.

\subsection{Non-linear Speech Evolution}
The KF utilizes an auto-regression (AR) model for speech evolution which is usually not adequate to represent the temporal characteristics of speech in reality. A better strategy would be using a network to learn the non-linear mappings from the clean speech signals (hidden variables in the KF) in previous frames to the clean speech in the current frame.

\begin{figure}[t]
  \centering
  \includegraphics[width=80mm]{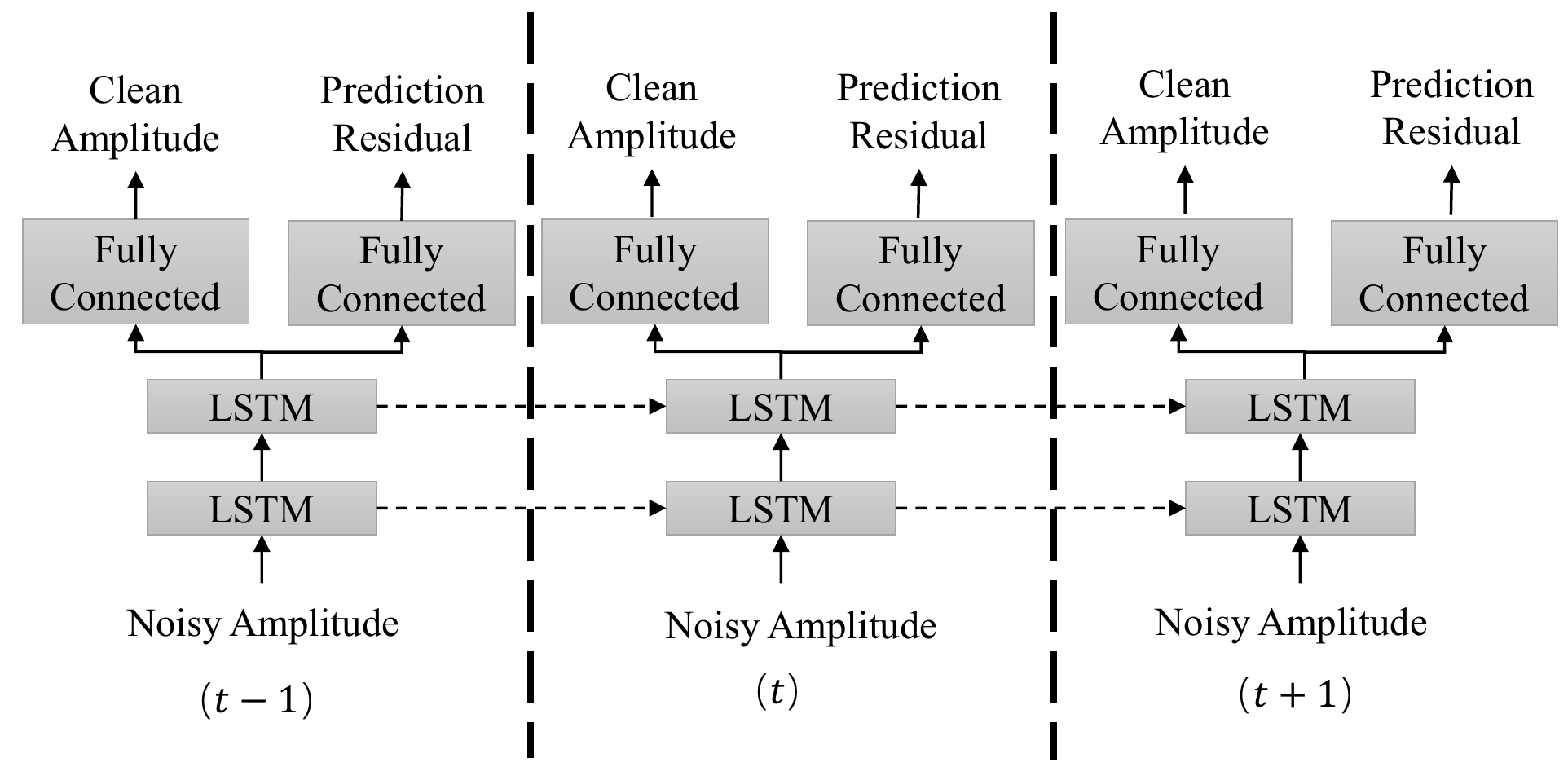}
  \caption{Structure of LSTM prediction network.}
  \label{fig:lstm}
\end{figure}

We note that the concept of hidden variable has been widely used by the RNN, which has shown a strong capability of sequential modelling and yield superior performances for the speech enhancement task \cite{Weninger2015,Gao2018,Sun2017}. Thus it provides a natural choice for the NKF to model the non-linear speech evolution. Here, a long short-term memory (LSTM) network is constructed as in Fig.~\ref{fig:lstm}, which learns two targets simultaneously. Similar to many conventional LSTM based speech enhancers, the network takes the noisy features as input, and predicts the clean amplitude of speech. Moreover, in accordance with KF, the prediction residual is also estimated, which will be used to determine the optimal NKF gain.

Specifically, in each frame, a feature vector is formed using the noisy amplitudes in all frequency bins, and a feature sequence can be obtained by stacking the feature vectors over a set of continuous frames. The feature sequence is first fed into the LSTM layers to model the temporal evolution of speech, and two separate fully-connected output layers are used to convert the LSTM outputs into the clean amplitude prediction and the prediction residual, respectively.

We note that unlike the KF that adopts a feedback loop to feed the KF estimation of the previous frame into the LP model, the LSTM uses the noisy amplitude spectrum as input. This is because we believe that the speech evolution has already been modelled by the hidden state propagation in the LSTM, and the additional noisy observation in each frame can help the LSTM to achieve more accurate predictions.

\subsection{Incorporating WF}
The speech amplitude estimated by the non-linear LSTM will be updated by incorporating the instantaneous observation in the current frame. In the conventional KF, by using the feedback loop to update the statistics of LP estimate and residual, it can be shown that incorporating the noisy input to update the KF output is equivalent to the WF \cite{Xue2018} when LP is excluded. Since the feedback loop is not used by the proposed NKF, in this section, based on the derivations in the KF, the WF is directly applied to incorporate the instantaneous information. The WF filters the signals only in the current frame, instead of performing prediction using several previous frames as is done by the LSTM.

Under the MMSE criterion, the optimal WF $H_{\textrm{Wiener}}(t,f)$ is given by
\begin{align}
H_{\textrm{Wiener}}(t,f) = \frac{\sigma^2_x(t,f)}{\sigma^2_y(t,f)} = 1-\frac{\sigma^2_v(t,f)}{\sigma^2_y(t,f)},
\label{wf0}
\end{align}
where $\sigma^2_y(t,f)$, $\sigma^2_x(t,f)$ and $\sigma^2_v(t,f)$ are the variances of the noisy speech, clean speech and additive noise, respectively. In practice, $\sigma^2_y(t,f)$ is computed from the noisy observation, and as shown in Fig.~\ref{fig:nkf}, a noise estimation network is constructed based on the simple ReLU-activated FNN. The network takes the noisy amplitudes and variances for frames within a left-side context window as input, and predicts the noise variances.
Then the WF output is obtained as
\begin{align}
|\hat{X}(t,f)|_{\textrm{Wiener}} = H_{\textrm{Wiener}}(t,f)|Y(t,f)|.
\label{wf1}
\end{align}

\subsection{Linear Weighting}
The clean amplitude estimates from LSTM and WF are finally combined by linear weighting to yield the NKF output $|\hat{X}(t,f)|_{\textrm{NKF}}$. With KF gain, we similarly define an NKF gain:
\begin{align}
G_{\textrm{NKF}} = \frac{\sigma^2_{r}(t,f)}{\sigma^2_{r}(t,f)+\sigma^2_{v}(t,f)},
\end{align}
where $\sigma^2_{r}(t,f)$ is the variance of LSTM prediction residual. Then the NKF output is calculated by:
\begin{align}
|\hat{X}|_{\textrm{NKF}} = G_{\textrm{NKF}}*|\hat{X}|_{\textrm{Wiener}}+(1-G_{\textrm{NKF}})*|\hat{X}|_{\textrm{LSTM}},
\end{align}
where $|\hat{X}|_{\textrm{LSTM}}$ denotes the LSTM estimation, the TF bin index ``$(t,f)$'' is omitted for simplicity.

Since the KF and WF are all derived under the MMSE criterion, the NKF network is optimized by taking the MSE between the clean amplitude and the NKF estimation as the loss function. The time-domain speech is recovered by inverse STFT which uses the phase of the noisy speech.

\section{Experiment}

\begin{figure}[t]
  \centering
  \includegraphics[width=80mm]{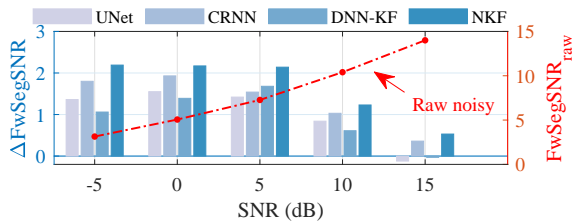}
  \caption{Improvements of FwSegSNR for different SNRs.}
  \label{fig:snr}
\end{figure}

\subsection{Experimental Setup}
We use the clean speech database, Librispeech \cite{Panayotov2015}, and two noise databases including PNL-100Nonspeech-Sounds (PNL) \cite{Hu2010a} and the noise subset of MUSAN corpus (MUSAN-Noise) \cite{musan2015} to prepare the training and testing data.

Unmatched noisy conditions for training and testing are generated. A 100-hour training set and a 10-hour development set are obtained by mixing the ``CLEAN-360'' subset of Librispeech and the MUSAN-Noise, under a speech-to-noise ratio (SNR) randomly chosen from $\{-6:3:21\}$~dB. The ``TEST-CLEAN'' subset of Librispeech and the PNL noise set are used to produce a 10-hour test set at SNR levels of  $\{-5:5:15\}$ dB. The sample rate of all signals is $16$~kHz, and the analysis window for STFT and feature extraction is $256$ samples with a $75\%$ overlap.

For the proposed NKF, the LSTM prediction network has two 1024-node hidden LSTM layers, and the noise estimation FNN has three layers with 1024 nodes in the hidden layer. The left-side context window for the noise estimation FNN is $30$ and $\sigma^2_y(t,f)$ in \eqref{wf0} is computed using previous $20$ frames.

Three baseline methods working in the TF domain are used for comparison, and they include a) an UNet which is based on CNN and uses the encoder-decoder framework with skip connections; b) a CRNN which adopts an LSTM into the UNet to capture the temporal characteristics; c) a DNN-KF which follows \cite{Yu2019} to resolve the LP coefficient and noise estimation problem by the neural networks, and uses the conventional KF for speech enhancement. During training, the batch size of all methods is 16, and the sequence length is 2048 frames. The networks are trained by 20 epochs.

Three objective speech quality measures including the frequency-weighted segmental SNR (FwSegSNR),  perceptual evaluation of speech quality (PESQ)  and short-Time Objective Intelligibility (STOI) \cite{Loizou2013} are used for evaluation. For all metrics, higher value indicates better performance.

\begin{figure}[t]
  \centering
  \includegraphics[width=80mm]{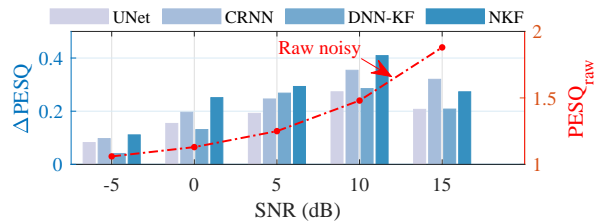}
  \caption{Improvements of PESQ for different SNRs.}
  \label{fig:pesq}
\end{figure}

\begin{figure}[t]
  \centering
  \includegraphics[width=80mm]{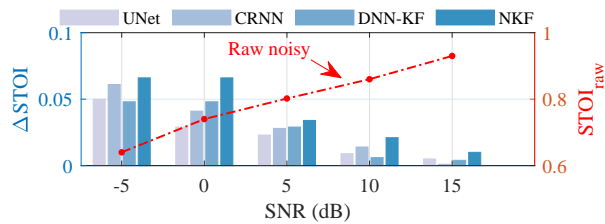}
  \caption{Improvements of STOI for different SNRs.}
  \label{fig:stoi}
\end{figure}

\subsection{Results}

Results of different methods for different SNRs are depicted from Fig.~\ref{fig:snr} to Fig.~\ref{fig:stoi}. In  Fig.~\ref{fig:snr}, it can be observed that the proposed NKF can consistently yield the highest improvement on the FwSegSNR over the noisy speech, which indicates the effectiveness of the NKF to suppress noise. The NKF also has the best capability to preserve speech when performing noise reduction, which is shown in Fig.~\ref{fig:pesq} and Fig.~\ref{fig:stoi} that the highest improvements of PESQ and STOI are achieved.

 We can see that generally different methods yield the most significant improvements when the SNR is 0~dB and 5~dB. The speech features would be highly contaminated by noise when the noise level is extremely high, making the speech enhancement problem more difficult. On the other hand, since the objective measures for speech in high SNR cases are already high, there is less potential for improvement. It is also worth noting that the DNN-KF which relies on an external processor to estimate the LP coefficients and the noise level achieves the worst performances in low SNR conditions, which is due to that the inaccurate information provided by the external processor, and that the assumption of linear speech evolution model  is not satisfied.

\section{Conclusion}
An NKF based speech enhancement method  is proposed by integrating the DNN with the statistical signal processing, following the framework of conventional KF. The statistical signal processing components can be seen as providing priori expert knowledge and serve as a regularization for the network. Experimental results in different noisy conditions show the effectiveness of the proposed method.

\small
\bibliographystyle{IEEEtran}
\bibliography{IEEEabrv,sapref}

\end{document}